\documentclass[reprint,amsmath,amssymb,aps,prl,superscriptaddress,longbibliography,floatfix,groupedaddress]{revtex4-2}
\usepackage{graphicx}
\usepackage[utf8]{inputenc}
\usepackage{soul}
\usepackage[usenames]{xcolor}
\usepackage{dcolumn}
\usepackage{bm}
\usepackage{textgreek}  
\usepackage{amsmath}

\newcommand{\ghop}{g_\mathrm{hop}}

\newcommand{\Phop}{P_{\mathrm{hop}}}
\newcommand{\TMCT}{T_\mathrm{MCT}}

\begin{document}

\title{String-like Rearrangements Induce Mobility in Supercooled Liquids}

\author{Rahul N. Chacko}
\affiliation{Department of Physics and Astronomy, University of Pennsylvania, Philadelphia, Pennsylvania 19104, USA}

\date{\today}

\begin{abstract}
Dynamical heterogeneity is a signature phenomenon of deeply supercooled liquids and glasses.
Here, we demonstrate that the spatiotemporal correlations between local relaxation events that underpin it are the result of
local relaxation events raising the likelihood that other relaxation events subsequently occur nearby,
confirming a widely held, but until now unsubstantiated,
belief that dynamical facilitation is responsible for dynamical heterogeneity.
We find that mobility is propagated through the entrainment of particles into elementary string-like rearrangements,
known as microstrings, and not through perturbing the structure surrounding these rearrangements.
\end{abstract}

\maketitle

Mobility in deeply supercooled liquids is not uniformly distributed in space~\cite{edigerSpatiallyHeterogeneousDynamics2000},
but is instead concentrated into clusters of high mobility that grow as a function of the time scale of observation%
~\cite{candelierSpatiotemporalHierarchyRelaxation2010, keysExcitationsAreLocalized2011, scallietThirtyMillisecondsLife2022}.
This phenomenon, known as dynamical heterogeneity,
is thought to explain many of the signature features of supercooled liquids and glasses,
from non-exponential relaxation processes~\cite{kobDynamicalHeterogeneitiesSupercooled1997, bohmerNatureNonexponentialPrimary1998}
to the breakdown of the Stokes-Einstein relation%
~\cite{hodgdonStokesEinsteinViolationGlassforming1993, tarjusBreakdownStokesEinstein1995, senguptaBreakdownStokesEinsteinRelation2013}
to excess wings in relaxation spectra~\cite{guiselinMicroscopicOriginExcess2022, scallietThirtyMillisecondsLife2022}.
However, the origin of dynamical heterogeneity remains an open question%
~\cite{wolynesSpatiotemporalStructuresAging2009, chackoElastoplasticityMediatesDynamical2021, scallietThirtyMillisecondsLife2022,
ozawaElasticityFacilitationDynamic2023, tahaeiScalingDescriptionDynamical2023, hasyimEmergentFacilitationGlassy2024},
with debate surrounding the respective roles of structure and dynamics in this origin%
~\cite{edigerSpatiallyHeterogeneousDynamics2000, widmer-cooperHowReproducibleAre2004, widmer-cooperRelationshipStructureDynamics2005,
widmer-cooperPredictingLongTimeDynamic2006, berthierStructureDynamicsGlass2007, wisitsorasakDynamicalHeterogeneityGlassy2014, gavazzoniTestingTheoriesGlass2024}
evoking the debate between thermodynamic and dynamical explanations of the glass transition%
~\cite{tarjusOverviewTheoriesGlass2011, wyartDoesGrowingStatic2017, biroliRFOTTheoryGlasses2024, gavazzoniTestingTheoriesGlass2024}.

At the local scale, it is clear that dynamics in a deeply supercooled liquid is influenced by structure.
Molecular dynamics studies in the isoconfigurational ensemble~\cite{widmer-cooperHowReproducibleAre2004},
comprising equilibrium trajectories with the same initial configuration but with different initial velocity sets,
find that a few rare particles have an exceptionally high propensity to hop out of the cage formed by their nearest neighbors
and into a new cage%
~\cite{widmer-cooperHowReproducibleAre2004, widmer-cooperPredictingLongTimeDynamic2006, rodriguezfrisTimeEvolutionDynamic2009}.
Correspondingly, a wide array of local structural indicators of mobility%
~\cite{richardPredictingPlasticityDisordered2020} have been identified.
However, the length scales of short-time propensity and of the local structural indicators currently known to us
are much shorter than that of the dynamical heterogeneity seen on long time scales%
~\cite{widmer-cooperPredictingLongTimeDynamic2006, tongStructuralOrderGenuine2019, lerbingerRelevanceShearTransformations2022}.

Tracking the locations of cage hops shows them to be correlated in space and time%
~\cite{candelierSpatiotemporalHierarchyRelaxation2010, keysExcitationsAreLocalized2011, chackoElastoplasticityMediatesDynamical2021,
scallietThirtyMillisecondsLife2022},
suggesting an explanation for dynamical heterogeneity as the cumulative effect of this correlation.
In this case, the question of structure versus dynamics becomes a question of correlation versus causation:
To what extent are cage hops correlated merely because they reside within spatially correlated weak structure,
rather than because one hop causes subsequent hops to occur nearby?

In the causal scenario, known as dynamical facilitation~\cite{chackoDynamicalFacilitationGoverns2024},
dynamical heterogeneity is a fundamentally dynamical phenomenon.
Models of dynamical facilitation range from those in which structural correlations play no role,
as in kinetically constrained models (KCMs)~\cite{ritortGlassyDynamicsKinetically2003},
to others in which hopping events create spatially correlated regions of weak structure at their location%
~\cite{ozawaElasticityFacilitationDynamic2023, tahaeiScalingDescriptionDynamical2023,
hasyimEmergentFacilitationGlassy2024, ridoutDynamicsMachinelearnedSoftness2024},
as in elastic models~\cite{ozawaElasticityFacilitationDynamic2023, tahaeiScalingDescriptionDynamical2023, 
hasyimEmergentFacilitationGlassy2024},
showing that dynamical heterogeneity can result from very different relationships between structure and dynamics.
Identifying the origin of dynamical heterogeneity entails identifying the particular interplay
between structure and dynamics that is responsible for it in supercooled liquids and structural glasses.

In this paper we investigate the origin of dynamical heterogeneity in molecular dynamics simulations of a model glass former.
We first identify a simple, physically motivated characterisation of cage hops and demonstrate that
dynamical heterogeneity indeed results from the spatiotemporal correlation of hopping events.
We then exploit the isoconfigurational ensemble to show that this correlation reflects an underlying causal relationship between events.
Further exploitation of this ensemble allows us to identify the mechanism underlying mobility propagation.
In supercooled liquids and glasses, rearrangement events occur as ``microstrings'' of collectively hopping particles%
~\cite{donatiStringlikeCooperativeMotion1998, gebremichaelParticleDynamicsDevelopment2004}.
We find that direct involvement of a particle in such a rearrangement event increases mobility on average,
whereas mere adjacency to a rearrangement does not promote mobility.
Direct involvement in a rearrangement microstring is thus the primary mode by which mobility propagates itself,
a result consistent with dynamical facilitation as mediated by localised  excitations%
~\cite{garrahanCoarsegrainedMicroscopicModel2003, keysExcitationsAreLocalized2011}.

We simulate in two dimensions a model supercooled liquid composed of particles $i$ of identical mass $m$
and sizes $\sigma_i$ randomly sampled from the interval $\left[ \sigma_\mathrm{min}, \sigma_\mathrm{max} \right]$
according to the distribution $P \left( \sigma \right) \propto \sigma^{-3}$.
The size limits $\sigma_\mathrm{min}$ and $\sigma_\mathrm{max}$ are chosen such that they enforce the expected value $\bar \sigma = 1$
and coefficient of variation $c_\sigma = 0.23$ of this distribution.
Pairs of particles $i$ and $j$ with center-to-center separation $r$ and with normalised separation
$\tilde r := r / \left[ \frac{1}{2} \left( \sigma_i + \sigma_j \right) \left( 1 - 0.2 \left| \sigma_i - \sigma_j \right| \right)\right]$
below the interaction cutoff $\tilde r_\mathrm{cut} = 1.25$
interact via the pair potential $V \left( \tilde r \right) = V_0 \left[ \tilde r^{-12} + c_0 + c_2 \tilde r^2 + c_4 \tilde r^4 \right]$,
where the constant coefficients $c_0$, $c_2$, and $c_4$ are chosen such that
$V \left( \tilde r_\mathrm{cut} \right)$, $V^\prime \left( \tilde r_\mathrm{cut} \right)$, and $V^{\prime \prime} \left( \tilde r_\mathrm{cut} \right)$
all vanish.
We non-dimensionalise by choosing $m$, $\bar \sigma$, $V_0$ and $V_0 / k_\mathrm{B}$
to respectively be our mass, distance, energy, and temperature units, where $k_\mathrm{B}$ is the Boltzmann constant.

This system, introduced in \cite{ninarelloModelsAlgorithmsNext2017},
can by design be efficiently equilibrated to low temperatures using a swap Monte Carlo algorithm.
In this way we obtain equilibrium configurations of $N=10000$ particles
contained in a fully periodic square box of side length $L=100$
at temperature $T = 0.110$ below the mode coupling theory temperature $\TMCT \approx 0.130$~\cite{SM}.
We use these as initial configurations from which to generate isoconfigurational ensembles of equilibrium trajectories,
with each trajectory corresponding to a microcanonical molecular dynamics simulation.
When analysing these trajectories, we quench each snapshot of the system to the local inherent state to
suppress noise from thermal fluctuations within cages~\cite{keysExcitationsAreLocalized2011}.
We account for Mermin-Wagner fluctuations in this two-dimensional system~\cite{flennerFundamentalDifferencesGlassy2015}
by using cage-relative displacements%
~\cite{mazoyerDynamicsParticlesCages2009, vivekLongwavelengthFluctuationsGlass2017, illingMerminWagnerFluctuations2017}
with cage radius 8.7~\cite{SM}.

\begin{figure}
\includegraphics[width=\columnwidth]{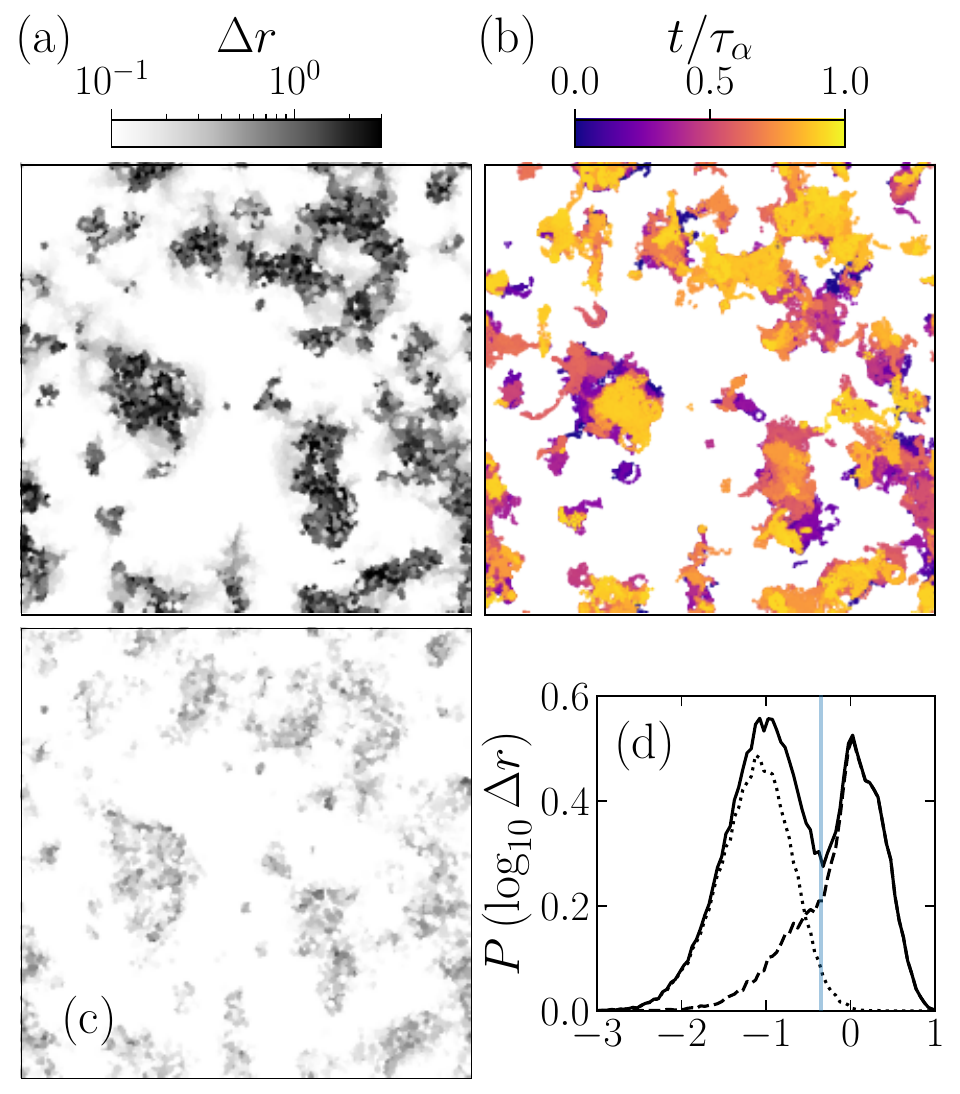}
\caption{Particle mobility at temperature $T=0.110$ across $\tau = 2 \times 10^5$ time units.
(a): Snapshot of particle displacements $\Delta r (\tau)$.
(b): Locations of cage hops in interval $[t, t+\Delta t]$ for $t=0$, $\Delta t$, $2 \Delta t$, $\dots$, $2000 \Delta t$
(darker to lighter), where $\Delta t = 10^2$.
(c): Snapshot of isoconfigurationally averaged particle displacements $\Delta r_\mathrm{iso} (\tau)$.
(d): Solid: Distribution of $\log_{10} \Delta r (\tau)$.
Dotted: Component of this distribution due to particles not seen to hop out of their cage within any of the $\Delta t$-sized intervals.
Dashed: Component attributable to particles seen to hop at least once.
The transparent vertical line corresponds to the cage hop threshold $\Delta r^2 = 0.2$.
\label{fig:hops_vs_dr}}
\end{figure}

In Fig.~\ref{fig:hops_vs_dr}, we show the equilibrium dynamics of this system
across $2.00 \times 10^5$ units of time (c.f. the bulk relaxation time scale $\tau_\alpha = 2.17 \times 10^5$ \cite{SM}).
Heterogeneous dynamics can be seen in Fig.~\ref{fig:hops_vs_dr}(a),
with clusters of mobile particles visible in a sea of immobile particles.
The distribution of displacements, shown in Fig.~\ref{fig:hops_vs_dr}(d), is bimodal,
as expected for a supercooled liquid below $\TMCT$%
~\cite{sastrySignaturesDistinctDynamical1998, berthierStructureDynamicsGlass2007}.
Mobile and immobile particles can therefore be considered distinct particle populations,
with the temperature-independent~\cite{SM} position of the minimum separating the two modes providing a criterion for identifying cage hops:
We consider a particle $i$ to hop out of its cage in interval $I = [t, t + \Delta t]$
if between $t$ and a short time $t + \Delta t$ later, it has squared displacement $\Delta r^2 > 0.2$.

In Fig.~\ref{fig:hops_vs_dr}(b), we show the positions of cage hops in intervals $[t, t + \Delta t]$ 
for $t=0$, $\Delta t$, $2 \Delta t$, $\dots$, $2000 \Delta t$, where $\Delta t = 10^2$.
A good correspondence between the locations of hopping events in Fig.~\ref{fig:hops_vs_dr}(b)
and the dark patches corresponding to mobile regions in Fig.~\ref{fig:hops_vs_dr}(a) can be seen.
This is confirmed by plotting separate displacement histograms for particles identified as hopping out of their cage
in at least one interval and particles that remain in their original cage throughout in Fig.~\ref{fig:hops_vs_dr}(d),
finding that almost the entire high mobility peak is accounted for by particles that hop.
Thus, if we can explain the correlation between hopping events in space and time,
we will know the origin of the clustering of mobility that characterises dynamical heterogeneity.

To this end, we investigate the process of mobility propagation,
beginning with the question of causation.
We study as a proxy for mobility propagation
the influence of a particle $i$ hopping out of its cage in interval $I_0 = [0, \Delta t]$
on the likelihood that a neighbouring particle $j$,
which did not hop in $I_0$, hops in the following interval $I_1 = [\Delta t,  2 \Delta t]$%
~\cite{vogelSpatiallyHeterogeneousDynamics2004, gokhaleGrowingDynamicalFacilitation2014}.
We choose $\Delta t \in \left\{ 3, 10, 30, 100 \right\}$.
Here, $\Delta t=3$ corresponds to the temperature-independent beginning of the caging plateau~\cite{SM},
and is therefore the smallest possible time scale over which all particles can sense their cage,
such that a displacement-based characterisation of caging is meaningful.
The scope for chains of multiple rearrangements
(``avalanches''~\cite{candelierSpatiotemporalHierarchyRelaxation2010}) to occur within $I_0$ or $I_1$
increases as $\Delta t$ is increased above $3$.

For each pair of particles $i$ and $j$ in a given configuration,
we quantify the extent to which $i$ hopping in $I_0$ causes $j$ to hop in $I_1$
via the dynamical facilitation ratio
\begin{equation}
    \phi_\mathrm{d.f.} = \left. \frac{P \left( j^1 | i^0 \right)}{P \left( j^1 \right)} \right|_{\textrm{not } j^0},
    \label{eq:phi_df}
\end{equation}
where $P \left( j^\alpha | i^\beta \right)$ denotes the conditional probability that particle $j$ hops in $I_\alpha$
given that particle $i$ hops in $I_\beta$,
and where $\left. \cdot \right|_{\textrm{not } j^0}$ indicates that the numerator and denominator of Eq.~\ref{eq:phi_df}
are calculated within the subset of isoconfigurational trajectories in which particle $j$ does not hop in $I_0$.
In essence, $\phi_\mathrm{d.f.}$ compares the hopping probability of the  formerly immobile particle $j$
in the wake of a hop by particle $i$ to the overall probability that $j$ hops.

\begin{figure}
\includegraphics[width=\columnwidth]{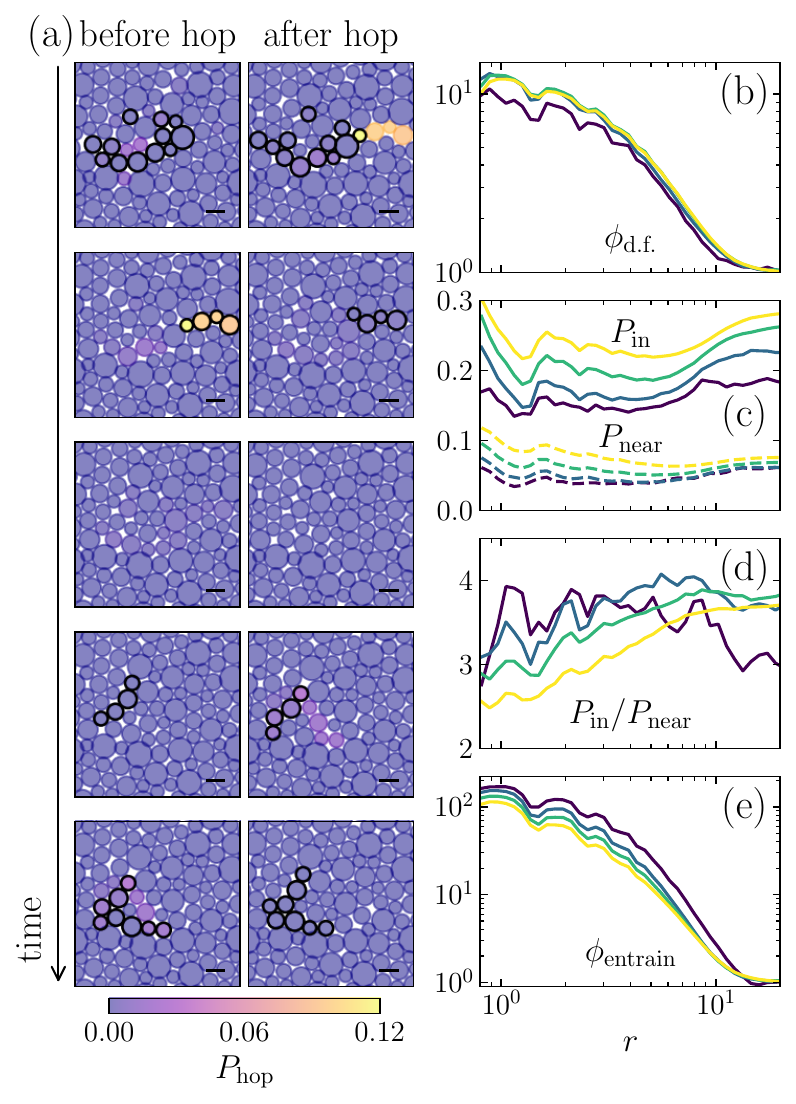}
\caption{(a): Snapshots of an equilibrium trajectory, spaced $100$ time units apart,
with the ``after hop'' snapshot on the $i$th row duplicated as the ``before hop'' snapshot on the $i+1$th row.
Particles identified as hopping out of their cage between the ``before hop'' and ``after hop'' snapshots are marked
with a thick black border.
Particles are coloured according to their isoconfigurational probability $\Phop$ of hopping within time $\Delta t=3$.
The scale bar indicates unit length.
(b): Dynamical facilitation ratio $\phi_\mathrm{d.f.}$, as defined in Eq.~\ref{eq:phi_df}.
(c)--(e): Comparison of the probability of hopping given prior involvement in a recent rearrangement event,
$P_\mathrm{in}$, to the probability of hopping given non-involvement in a recent rearrangement event a distance $r$ away,
$P_\mathrm{near}$ (see Eq.~\ref{eq:PinPnear}).
Curves correspond to interval lengths $\Delta t=3$, $10$, $30$ and $100$ (darker to lighter).
\label{fig:propagation}}
\end{figure}

We average $\phi_\mathrm{d.f.}$ over all pairs of particles $i$ and $j$ with given interparticle separation $r$
and over multiple independent configurations,
excluding pairs of particles with an undefined value of $\phi_\mathrm{d.f.}$ as measured from $400$ trajectories.
As shown in Fig.~\ref{fig:propagation}(b), we find that the relationship between $i$ hopping
and $j$ hopping is strongly causal,
with $\phi_\mathrm{d.f.} \approx 10$ at $r \approx 1$.
Increasing $\Delta t$ above $3$ to allow more time for avalanching leads to an increase in $\phi_\mathrm{d.f.}$,
suggesting that avalanches mediate this relationship.

Observation of cage hops in trajectories at low temperature reveals a surprising trend.
As illustrated in Fig.~\ref{fig:propagation}(a),
the microstrings of collectively hopping particles through which structural reorganisation proceeds%
~\cite{donatiStringlikeCooperativeMotion1998, gebremichaelParticleDynamicsDevelopment2004, keysExcitationsAreLocalized2011}
typically include one or more particles that recently hopped as part of an earlier microstring,
as opposed to merely residing in its neighbourhood.
To quantify this tendency, we introduce
\begin{equation}
\begin{aligned}
    P_\mathrm{in} &= \left. P \left( j^1 | j^0 \right) \right|_{i^0} \quad\quad \textrm{and} \\
    P_\mathrm{near} &= \left. P \left( j^1 | \textrm{not } j^0 \right) \right|_{i^0}.
\end{aligned}
\label{eq:PinPnear}
\end{equation}
At our sub-$\TMCT$ temperature, rearrangement events are rare, and it is reasonable to assume that
near-separated ($r \lesssim 2$, say) particles $i$ and $j$ that both hop in $I_0$ are part of the same microstring,
especially for short $\Delta t$.
In this case, $P_\mathrm{in}$ measures the probability that $j$ hops following participation in the same
microstring as $i$,
while $P_\mathrm{near}$ measures the probability that $j$ hops if it was near the microstring involving $i$,
but not part of it.

\begin{figure}
\includegraphics[width=\columnwidth]{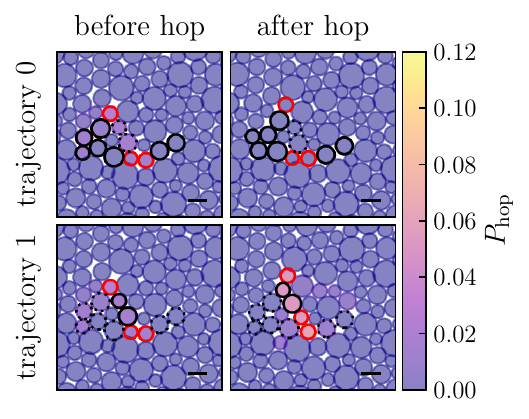}
\caption{Top row: trajectory corresponding to the bottom row of Fig.~\ref{fig:propagation}(a).
Bottom row: alternative equilibrium trajectory with the same initial configuration as the top row.
Particles that hop in a given trajectory are indicated with thick black outlines,
particles that hop in both trajectories are indicated with red outlines
and particles that do not hop in the given trajectory, but hop in the other trajectory,
are indicated with dotted black outlines.
\label{fig:localised_excitation}}
\end{figure}

We average $P_\mathrm{in}$ and $P_\mathrm{near}$ in the same manner as $\phi_\mathrm{d.f.}$,
while ensuring that we are comparing different sets of trajectories
and not different sets of particles (see Fig.~\ref{fig:localised_excitation})
by excluding from both averages any pair of particles for which either $P_\mathrm{in}$ or $P_\mathrm{near}$ is undefined.
For $\Delta t = 3$, we find (Fig.~\ref{fig:propagation}(c)) that neither $P_\mathrm{in}$ nor $P_\mathrm{near}$ is very sensitive to $r$,
aside from the undershoot near $r \approx 6$ (discussed in the End Matter).
For $P_\mathrm{in}$, this reflects the fact that a particle $j$ that hops in $I_0$
is part of some rearrangement event, even if not the same event as $i$.
For $P_\mathrm{near}$, this insensitivity to $r$ is more remarkable,
as it implies that mere proximity to a rearrangement is insufficient for the direct propagation of mobility.
As $\Delta t$ is increased to allow more time for avalanching, $P_\mathrm{near}$ develops a growing decay as a function of $r$,
indicating that it is avalanches that allow mobility to propagate to the neighbourhoods of rearrangements.
Finally, we note that for $\Delta t = 3$,
$P_\mathrm{in}/P_\mathrm{near} \approx 3.5$ for $r \lesssim 10$ (Fig.~\ref{fig:propagation}(d)),
demonstrating that participation in a rearrangement does on average correspond to an increase in mobility.

These results imply a scenario in which low-mobility particles are required to hop
in order to increase their (isoconfigurational) mobility.
We thus need microstrings to be able to entrain otherwise-immobile particles into hopping out of their cages.
We can assess this capability via the entrainment ratio
\begin{equation}
    \phi_\textrm{entrain} = \frac{P \left( j^0 | i^0 \right)}{P \left( j^0 \right)},
\label{eq:phi_entrain}
\end{equation}
which measures the extent to which pairs of particles $i$ and $j$ are entrained into moving together in $I_0$
as opposed to being independently mobile.
As we show in Fig.~\ref{fig:propagation}(e),
after averaging $\phi_\textrm{entrain}$ in the same manner as $\phi_\textrm{d.f.}$,
$\phi_\textrm{entrain} > 10^2$ up to $r \lesssim 3$ for $\Delta t = 3$.
Furthermore, increasing $\Delta t$ leads to a decrease in $\phi_\textrm{entrain}$,
indicating that this relationship between $i$ hopping and $j$ hopping is direct,
weakened by avalanching.
We conclude that otherwise-immobile particles can indeed be entrained into hopping.

In this paper, we have shown that dynamical heterogeneity
results from a causal relationship between hopping events and subsequent hopping events nearby.
Furthermore, we have shown that on average,
direct involvement in a rearrangement event increases particle mobility,
and that otherwise-immobile particles can be entrained into such events.
Together, these two results comprise a mobility propagation mechanism capable of explaining
the causal relationship between spatio-temporally close hopping events,
with entrainment forcing otherwise-immobile particles to sample new local environments.
By contrast, we have found that in spite of
mild elastic displacements~\cite{chackoElastoplasticityMediatesDynamical2021},
mere proximity to a rearrangement event does not directly lead to an increase in mobility,
as evidenced by the fact that for small $\Delta t$,
$P_\textrm{near}$ at small $r$ is no greater than $P_\textrm{near}$ at large $r$ (Fig.~\ref{fig:propagation}(c)).
Instead, the $\Delta t$ dependence of $P_\textrm{near}$ and $\phi_\textrm{d.f.}$ indicates that the propagation of mobility
from particles directly involved in a rearrangement event to particles nearby is mediated by chains of facilitated rearrangements.

These observations clarify the connection between local excitations and microstrings in supercooled liquids%
~\cite{vogelSpatiallyHeterogeneousDynamics2004, keysExcitationsAreLocalized2011, gokhaleGrowingDynamicalFacilitation2014,
ortliebProbingExcitationsCooperatively2023, chenVisualizingSlowInternal2023},
revealing a mechanism remarkably similar to the mechanism for dynamical facilitation in plaquette models%
~\cite{garrahanGlassinessEmergenceEffective2002, chackoDynamicalFacilitationGoverns2024}.
Indeed, when we apply our analysis to one such model,
we obtain results consistent with Fig.~\ref{fig:propagation} (see Fig.~\ref{fig:TPM_propagation}).
In plaquette models, excited plaquettes facilitate multiple mutually exclusive rearrangement events
and play an essential role in propagating mobility.
In our model supercooled liquid, we similarly find particles able to participate in multiple rearrangement events,
depending on the trajectory (Fig.~\ref{fig:localised_excitation}).
Due to time reversibility,
this class notably includes any particle involved in two rearrangements in quick succession
(hence propagating mobility) during the inter-rearrangement period.


In recent work, Ridout and Liu~\cite{ridoutDynamicsMachinelearnedSoftness2024}
present a lattice model of supercooled liquids designed around the constraint of detailed balance.
They find that perturbations induced by single-point rearrangements significantly under-predict dynamical heterogeneity,
but that multi-point rearrangements that strongly perturb the mobility at the rearranging points do suffice.
In light of our results here, it may be worth investigating if multi-point rearrangements alone,
neglecting long-range perturbations, suffice.

Our result that rearrangements do not directly enhance the mobility of particles in their neighbourhood
challenges the hypothesis that dynamical facilitation in supercooled liquids is mediated by
the elastic response of supercooled liquids to rearrangement events.
In previous work~\cite{chackoElastoplasticityMediatesDynamical2021},
this author presented evidence from simulations of an elastic signature in the positions of facilitated hopping events.
The results from this present study argue for a re-interpretation of these earlier results:
the elastic signature found there relates to individual microstrings,
not the correlation between distinct rearrangement events
(see End Matter).

Finally, in what concerns the relationship between structure and dynamics,
it is important to note that while we have shown that the correlation between cage hops reflects a causal relationship,
the microstrings through which mobility propagates are necessarily the result of kinetic constraints imposed by the local structure.
From the good correspondence between the particle displacement and isoconfigurational displacement in Fig.~\ref{fig:hops_vs_dr},
we can conclude that these structurally imposed constraints strongly restrict the dynamics, even on a $\tau_\alpha$ time scale:
the influence of the initial structure can evidently persist over time scales exceeding the relaxation time of the system.
This is easier to understand in a scenario in which mobility propagation is restricted to microstrings,
which are determined by the local structure,
than if mobility could propagate to any particle neighbouring a microstring.

A full description of dynamical facilitation in supercooled liquids will require further investigation into the role of structure.
In particular, the relationship between localised excitations and microstrings is central to the conclusions of this paper,
but neither is easy to characterise.
Microstrings obfuscate the former; a particle with high hopping probability
might represent a localised excitation but might equally be merely entrained in a microstring facilitated
by local structure elsewhere.
While much progress has been made towards predicting single-particle hopping probabilities%
~\cite{richardPredictingPlasticityDisordered2020}
or propensities~\cite{bapstUnveilingPredictivePower2020, boattiniAveragingLocalStructure2021, shibaBOTANBOndTArgeting2023, jungPredictingDynamicHeterogeneity2023, jungDynamicHeterogeneityExperimental2024, pezzicoliRotationequivariantGraphNeural2024},
the identification of structural precursors to microstrings as multi-particle objects remains an open problem.

R.N.C. thanks S. Arzash, L. Berthier, G. Biroli, O. Dauchot, R.C. Dennis, J. Kurchan, F.P. Landes, A.J. Liu, M. Ozawa, D.R. Reichman,
S.A. Ridout, A. Winn and M. Wyart for helpful discussions concerning this project.
A great debt of gratitude is owed to M.E. Cates, R.C. Dennis, S.M. Fielding and A.J. Liu
for their feedback on this manuscript.
This work was supported by the Simons Foundation via the ``Cracking the glass problem'' collaboration (Grant No.~454945)
and via a Simons Investigator grant (Grant No.~327939) and by NIH (U01-CA-254886).
R.N.C. thanks M. Wyart for hosting him at EPFL during the earliest stages of this project.
This work used the Anvil Supercomputer at the Rosen Center for Advanced Computing~\cite{songAnvilSystemArchitecture2022}
through Allocation No.~PHY200101 from the Extreme Science and Engineering Discovery Environment%
~\cite{townsXSEDEAcceleratingScientific2014},
which was supported by National Science Foundation Grant No.~1548562.

\appendix

\section{End Matter}

\emph{Statistics}---Our results in the main text are averaged over $1000$
independent initial configurations.
For each configuration, we generate an isoconfigurational ensemble of $400$
trajectories.
The isoconfigurational probabilities reported in the main text
are estimates based on the fraction of trajectories in which the random event in question is seen to occur.

\emph{Plaquette Model}---In the main text, we mention that the mechanism for mobility propagation
that we observe in our supercooled liquid strongly resembles dynamical facilitation as seen in plaquette models.
Here, we demonstrate this in detail for a specific plaquette model.

We consider the triangular plaquette model (TPM)
comprising $N=16384$ spins $s_{i, j} \in \left\{ -1, 1 \right\}$ on a fully periodic two-dimensional square lattice.
The indices $i, j \in \left\{ -64, \dots, 63 \right\}$ enumerate spins with unit spacing along the $x$ and $y$ axes, respectively.
We define plaquettes $p_{i, j} = s_{i, j} s_{i+1, j-1} s_{i+1, j}$.
In terms of these, the Hamiltonian is $H = - \frac{1}{2} \sum_{i, j} p_{i, j}$,
such that a rearrangement (spin flip) $s_{i, j} \mapsto - s_{i, j}$ corresponds to the simultaneous flips of
$p_{i-1, j}$, $p_{i-1, j+1}$ and $p_{i, j}$ and has corresponding energy barrier
$\Delta H_{ij} = p_{i-1, j} + p_{i-1, j+1} + p_{i, j}$.
We adopt Glauber dynamics~\cite{glauberTimeDependentStatistics1963}, such that at temperature $T$,
a spin $s_{i, j}$ is randomly selected and flipped with probability $1 / \left( 1 + \exp \left( \Delta H/T \right) \right)$
once every $1/N$ time units.
We fix $T = 0.35$.

Under these conditions, there is a one-to-one correspondence between spins and plaquettes~\cite{garrahanGlassinessConstrainedDynamics2000},
and it is in the plaquette representation that the analogy between our particulate system and the TPM can be seen.
The constraint that rearrangement events in the TPM correspond to triplets of plaquettes flipping together
is analogous to the constraint that particles hop out of their cages as part of microstrings.
The analogue of a mobile particle that can entrain others into a microstring is an excited plaquette $p_{i, j} = -1$,
since being excited will lower the barrier to rearrangement of all three spins $s_{i, j}$, $s_{i+1, j-1}$ and $s_{i+1, j}$
that involve plaquette $p_{i, j}$.

We define adjacent time intervals $I_0$ and $I_1$ as in the main text,
with rearrangement time scale $\Delta t \in \left\{ 10^{-2}, 10^{-1}, 10^0, 10^1, 10^2 \right\}$.
We consider a plaquette to have flipped in a given interval if it has different signs
at the beginning and end of the interval.
We calculate $\phi_\textrm{d.f.}$, $P_\textrm{in}$, $P_\textrm{near}$ and $\phi_\textrm{entrain}$
for the TPM as we did for the model supercooled liquid,
averaging over 1000 independent isoconfigurational ensembles of 1000 equilibrium trajectories.

\begin{figure}
\includegraphics[width=\columnwidth]{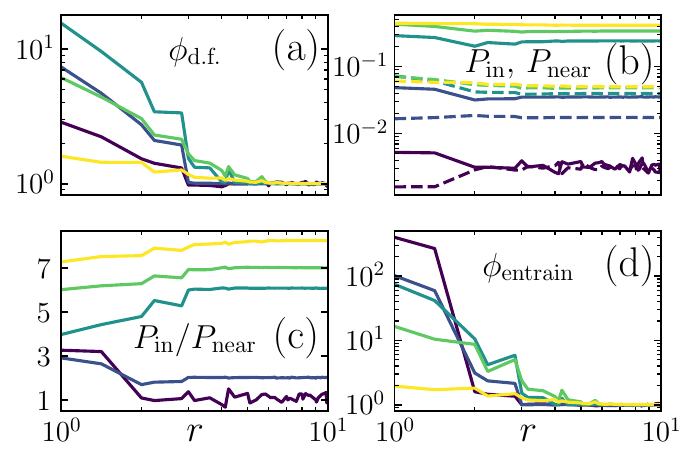}
\caption{(a): Dynamical facilitation ratio $\phi_\mathrm{d.f.}$, as defined in Eq.~\ref{eq:phi_df}.
(b): Comparison of (solid) the probability of flipping given prior
involvement in a recent rearrangement event, $P_\mathrm{in}$,
to (dashed) the probability of flipping given non-involvement
in a recent rearrangement event a distance $r$ away, $P_\mathrm{near}$ (see Eq.~\ref{eq:PinPnear}).
(c): $P_\mathrm{in} / P_\mathrm{near}$.
(d): Entrainment ratio $\phi_\mathrm{entrain}$, as defined in Eq.~\ref{eq:phi_entrain}.
Curves correspond to interval lengths $\Delta t=10^{-2}$, $10^{-1}$, $10^0$, $10^1$ and $10^2$ (darker to lighter).
\label{fig:TPM_propagation}}
\end{figure}

We see in Fig.~\ref{fig:TPM_propagation}(a) that $\phi_\textrm{d.f.} > 1$, confirming a causal link between plaquette flips,
as expected.
Notably, the only way for a rearrangement in $I_0$ to influence a plaquette not directly involved in it to flip in $I_1$
is via avalanches.
The $\Delta t$-dependence of $\phi_\textrm{d.f.}$ at small $r$,
which initially increases as more time is made available for avalanching
before decreasing at longer times as the structure relaxes,
is qualitatively similar to that of Fig.~\ref{fig:propagation}(a).

In Fig.~\ref{fig:TPM_propagation}(b), we see that $P_\textrm{in}$ is flat at large $r$ for the smallest $\Delta t$,
reflecting the fact that any flipping plaquette is part of some three-plaquette rearrangement,
but develops an undershoot at $r \approx 3$ as $\Delta t$ is increased.
This suggests that this undershoot, and by extension the undershoot in $P_\textrm{in}$
seen at $r \approx 6$ in Fig.~\ref{fig:propagation}(c), is a generic result of avalanching.

We also see that, similarly to $P_\mathrm{near}$ in Fig.~\ref{fig:propagation}(c),
$P_\mathrm{near}$ in Fig.~\ref{fig:TPM_propagation}(b) goes from having a positive slope at small $r$
to being flat and eventually developing an $r$-decay as $\Delta t$ is increased,
reflecting the fact that mobility in the TPM cannot directly propagate from one plaquette to another,
unless both flip together as part of the same rearrangement.

The trivial Hamiltonian means that at equilibrium, a fraction $1 / \left( 1 + \exp (1/T) \right)$ of plaquettes,
uniformly distributed throughout the system, are excited.
Maintaining this fraction implies equal rates of excitation and relaxation,
such that half of all plaquettes identified as having undergone a sign change across a given interval of time will be excited.
Given that few plaquettes are excited at low $T$, the average plaquette would see an increase in mobility upon flipping,
hence $P_\textrm{in}/P_\textrm{near} > 1$ as seen in Fig.~\ref{fig:TPM_propagation}(c).

Finally, we note that because entrainment into a triplet of flipping plaquettes is an instantaneous process,
increasing $\Delta t$ can only reduce $\phi_\textrm{entrain}$ as $\Delta t$ is increased and other rearrangement events
are allowed to occur.
This is indeed what we see in Fig.~\ref{fig:TPM_propagation}(d).

\emph{Elastoplastic Facilitation}---In previous work~\cite{chackoElastoplasticityMediatesDynamical2021},
this author studied the pair distribution of cage hops $g_\mathrm{hop} (r, \theta)$ in a similar glass former~\cite{SM},
where for each cage hop, $\theta=0$ is taken to lie along the extensional axis of the strain tensor local to the hop.
This quantity was found to be anisotropic, attaining maxima along the compressional and extensional axes of the strain tensor.
This was taken as evidence of an elastic signature of facilitation.

The results of the present study challenge this interpretation.
We find that facilitation is propagated through direct entrainment into microstrings,
rather than through the milder perturbations imposed by rearrangement events on their neighbourhoods.
Thus, a different interpretation of the earlier results suggests itself:
Rather than the anisotropy in $\ghop (r, \theta)$ reflecting stronger facilitation
along the compressional and extensional axes,
it may instead reflect the alignment and motion of individual microstrings.

\begin{figure}
\includegraphics[width=\columnwidth]{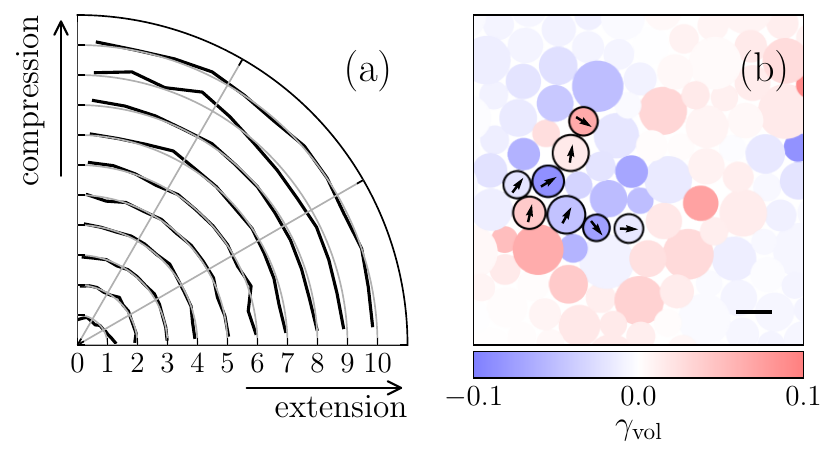}
\caption{Cage hops at temperature $T=0.110$ identified across time interval $\Delta t = 100$.
(a): $\ghop^\textrm{separate} (r, \theta)$ (see text)
at $r = 1$, $2$, $\dots$, $10$,
normalised by the mean value at the given $r$.
Each curve corresponds to a different $r$ value, with the fluctuations of the black curve
from the grey quarter-circle corresponding to fluctuations from the mean along $\theta$ at that $r$.
(b): System snapshot at the start of interval $I_0 = [0, \Delta t]$.
Particles identified as hopping out of their cage in $I_0$ are marked with a thick black border.
Particles are coloured according to their local volumetric strain $\gamma_\mathrm{vol}$ across $I_0$.
The scale bar indicates unit length.
\label{fig:elastoplasticity}}
\end{figure}

To test this new interpretation, we study the pair distribution
$\ghop^\textrm{separate} (r, \theta)$,
the normalised mean density of particles $j$ at position $\left( r, \theta \right)$
that hop out of their cage in interval $I_1$ but not $I_0$
given a particle $i$ at the origin that hops its cage in interval $I_0$ but not $I_1$.
This restriction ensures that the particles $i$ and $j$ are not part of the same microstring in either interval.
Here, $\theta=0$ corresponds to the extensional axis local to particle $i$ across $I_0$,
which will vary between trajectories.

As we show in Fig.~\ref{fig:elastoplasticity}(a) for intervals $I_0$ and $I_1$ of length $\Delta t = 100$,
$\ghop^\textrm{separate} (r, \theta)$
is isotropic at all but the shortest distances $r$.
For $r \geq 2$, fluctuations of this quantity along $\theta$ remain within $5\%$
of the mean.
Indeed, time reversibility at equilibrium implies symmetry under exchange of the
extensional and compressional axes
(except at small $r$, since the relative distances between particles will change across $I_0$),
so the mismatch between the $\theta \approx 0$
and $\theta \approx \pi/2$ values can be used to gauge the noise floor.
We see that fluctuations along $\theta$ do not systematically exceed this mismatch.

This confirms that for particles $j$ not directly involved in the same rearrangement as particle $i$, we do not see an elastoplastic signature.
Instead, the anisotropy seen in \cite{chackoElastoplasticityMediatesDynamical2021}
reflects individual microstrings aligning with Eshelby displacement quadrupoles~\cite{eshelbyDeterminationElasticField1957}.
The enhanced mobility of particles directly involved in a microstring
in turn explains why $\ghop$ remains anisotropic 
when restricting to pairs of particles
$i$ and $j$ that hop (non-exclusively) in separate time intervals
(see Supplemental Material of \cite{chackoElastoplasticityMediatesDynamical2021}).

In Fig.~\ref{fig:elastoplasticity}(b), we show the same snapshot as the bottom left plot of
Fig.~\ref{fig:propagation}(a),
with arrows indicating the extensional axes of the hopping particles and colours corresponding to the volumetric strain.
We see good alignment between the extensional axes of the hopping particles,
but no evidence of a volumetric strain dipole.

\bibliography{library,SMitem}

\end{document}